# Tunguska-1908 and similar events in light of the New Explosive Cosmogony of minor bodies

E.M.Drobyshevski

*Ioffe Physical-Technical Institute, Russian Academy of Sciences, 194021 St-Petersburg, Russia*
*E-mail: emdrob@mail.ioffe.ru*

*Abstract.* It is shown that the well-known Tunguska-1908 phenomenon (TP) problems (the fast transfer of the momentum and of the kinetic energy of the meteoroid $W \sim$ 10-50 Mt TNT to air, with its heating to $T > 10^4$ K at an altitude of 5÷10 km, the final turn of the smoothly sloping, $\delta \approx 0÷20°$ to horizon, trajectory of the body through ~10° to the West, the pattern and area of the tree-fall and trees' scorching by heat radiation, etc.) allow a fairly straightforward solution within the paradigm of New Explosive Cosmogony (NEC) of minor bodies, as opposed to other approaches. The NEC considers the short-period (SP) cometary nuclei, to which the Tunguska meteoroid belonged, to be fragments produced in explosions of massive icy envelopes of Ganymede-type bodies saturated by products of bulk electrolysis of ices to the form of a $2H_2+O_2$ solid solution. The nearly tangent entry into the Earth's atmosphere with $V \sim 20$ km/s of such a nucleus, ~200÷500 m in size and ~(5÷50)×$10^{12}$ g in mass, also saturated by $2H_2+O_2$, initiated detonation of its part with a mass of ~$10^{12}$ g at an altitude of 5÷10 km. This resulted in deflection of this fraction trajectory by 5°÷10°, and fast expansion with $V_t \approx 2$ km/s of the products of its detonation brought about their fast slowing down by the air, heating of the latter to $T > 10^4$ K and a phenomenon of moving high-altitude explosion, with the resultant scorching and fall of trees in a butterfly pattern. On crossing the Earth's atmosphere, the main part of the unexploded nucleus escaped into space, and this body moving presently in an SP orbit should eventually be identified in time. Its impact with $W \sim 250÷3000$ Mt TNT on the Earth's surface (which could occur in 1908, and now can be expected to happen in the future) would have produced a crater up to ~3.5÷8 km in size, with an ensuing ejection of dust that would have brought about a large-scale climatic catastrophe. The physical processes involved in the TP are pointed out to resemble those accompanying falling P/Shoemaker-Levy 9 fragments onto Jupiter and, possibly, the impact-caused Younger Dryas cooling on the Earth ~13 ka ago.

*The running title:* Tunguska-1908 and the New Explosive Cosmogony of comets

## 1. Introduction

More than one hundred years ago, on the morning of June 30, 1908, a bolide challenging the Sun in brightness crossed the sky from the SE to NW, starting its path from somewhere over the northern shore of lake Baikal or even the upper reaches of river Vitim. Its ~600÷800-km flight culminated in an explosion of energy $W \sim 10÷50$ Mt TNT (1 Mt TNT = 4.18×$10^{22}$ erg) at an altitude $H = 5÷10$ km over wild marshy taiga (60.9° N, 101.9° E), with a butterfly-shaped tree-fall over an area of ~2150 km$^2$ and radiation-scorched trees within ~200 km$^2$. The explosion initiated strong seismic and acoustic waves, perturbations in the geomagnetic field, and "light night skies" for the following three consecutive evenings within a latitude belt from 41° to 60° extending up to the Atlantic Ocean (Zotkin 1966).

If the meteoroid had fallen $4^h47'$ later, its explosion would have devastated St.-Petersburg (Astapovich 1934), leaving about one million people dead.

Copious literature was devoted to description of the Tunguska phenomenon (TP) and its analysis (about 600÷700 publications; see, e.g., Krinov (1949, 1963), Turco *et al* (1982),

Bronshten (2000), Vasilyev (1998, 2004), Longo (2007) and refs. therein); one could hardly say, however, that its mystery has been lifted, if for no other reason than the essential part played here by the sometimes fairly contradictory eyewitness accounts that are not always easy to reconcile with the existing scientific concepts.

## 2. Explosion energetics and altitude

The absence of an impact crater and the tremendous forest devastation imply unambiguously the above-the-ground, atmospheric character of the explosion. Astapovich (1933) estimated originally its energy as $\sim 4.4 \times 10^{21}$ erg from the forest fall, and $10^{23} \div 10^{24}$ erg, from seismic data. With the advent of the era of atmospheric nuclear explosions, these controlled seismically and acoustically experiments yielded an estimate of $W \sim 10 \div 50$ Mt TNT for the energy, and $5 \div 10$ km for the altitude of the explosion (see the summary in Ben-Menachem 1975, Bronshten 2000, Vasilyev 1998, 2004, Longo 2007). Whence it follows that assuming a velocity $V = 20 \div 30$ km/s, the mass of the object could be $\sim 10^{12}$ g. One should naturally treat with care all these estimates; indeed, atomic explosions are nothing else than only an approximation to a fast moving and certainly not point-type TP blast.

It is worthwhile to note that the fall of 20+ fragments of comet P/Shoemaker-Levi 9 (SL-9) on Jupiter in 1994 with $W \sim 10^5 \div 10^6$ Mt TNT (e.g. Klumov *et al* 1994, Crawford *et al* 1995, Boslough and Crawford 1997) seems to approach most closely in its associated physics to the TP. Another comet nucleus fall resembling the TP occurred, as it is supposed by Firestone *et al* (2007), in the North America ~12.9 ka ago. It resulted in climatic catastrophe in the Northern hemisphere which caused the Younger Dryas cooling and the megafaunal extinction (Kennett *et al* 2008). Study of these falls stresses the significance of detailed studies of the recent TP.

## 3. The origin of the TP object: on the possibility of its kinetic energy transfer to the air

It is presently universally accepted that the TP was initiated by the fall of an asteroid or a nucleus of a small comet, $\sim 50 \div 100$ m in size. The absence of fragments thereof (see, however, below) favors rather the icy cometary nature of the object (Zotkin 1969, Kresák 1978, Asher and Steel 1998), although trajectory-based considerations make the asteroid hypothesis more plausible (Sekanina 1983, 1998, Farinella *et al.* 2001, Jopek *et al* 2008).

We come now to a key question of how to convert the kinetic energy of an object moving with $V = 20 \div 30$ km/s into the energy of explosion at an altitude $H = 5 \div 10$ km, where the air density $\rho_a$ is one half to one third the ground level value. For this to be possible, the air has to remove the energy from the body by braking it so fast that its temperature remains above $T > 10^4$ K (the air temperature behind the shock wave with $V_s = 20$ km/s at an altitude $H = 8$ km is $\sim 4 \times 10^4$ K, and $T \approx 2.4 \times 10^4$ K and $1.9 \times 10^4$ K at 50 and 100 km, respectively (Kuznetsov 1965)). This is possible if the body meets in its path in hypersonic motion a mass of air larger than its own mass. As of today, a possible appropriate scenario appears to be the so-called "explosion in flight" (EF), when the body disintegrates under the velocity head $\rho_a V^2/2$ ($\approx 10^9$ dyne/cm$^2$ for $V = 20$ km/s at $H = 8$ km) which exceeds its strength. This results in a fast expansion of the ensemble of fragments due to their entrainment by the air flowing about the body, with the attendant sharp growth of the effective aerodynamic cross section of the ensemble.

Similar processes were discussed by many authors (e.g., Pokrovskii 1966, Fadeenko 1969, Grigoryan 1979, Shurshalov 1984, Chyba *et al* 1993, Hills and Goda 1993, Crawford *et al* 1995, Tirskiy and Khanukaeva 2008). It is assumed that the cloud of fragments expands initially (before their evaporation) in the transverse direction with a velocity $V_t \approx V(\rho_a/\rho)^{1/2}$ (e.g., Svetsov *et al* 1995), thus resulting in a fairly fast growth of the gas-dynamic cross section (for $V = 20$ km/s and $\rho = 2.5$ g/cm$^3$ - stone, $V_t \approx 300$ m/s; for $\rho = 1$ g/cm$^3$ - ice, $V_t \approx 450$ m/s). It is further assumed that hot air penetrates among the fragments and evaporates them completely, an aspect that should account for the absence of the Tunguska meteoroid fragments on the ground.

### 3.1. "Explosion in flight" (EF) and the relevant problems

There are, however, some aspects of the EF scenario that are difficult indeed to agree with on the spot.

*First*, the strength of a stone is $\sigma \sim (2.5 \div 8.3) \times 10^9$ dyne/cm$^2$ (Beresnev and Trushin 1976); nobody knows what is the bulk crack content of a real stone asteroid, and accepting the strengths of the meteorites collected on the ground as a reference point would be a hardly justifiable approach. The elastic limit for granite is $\approx 10^9$ dyne/cm$^2$, for solid water ice at $T = 257$ K it is $0.17 \times 10^9$ dyne/cm$^2$, while at $T = 81$ K it increases twice - $0.34 \times 10^9$ dyne/cm$^2$ (Lange and Ahrens 1987). The strength value is, correspondingly, higher by a factor of ~1.5-2. On the other hand, any person living in high latitude regions knows that cracks in ice have a trend to heal in time. *Second*, one seems to disregard possible strong (up to ~$10^2$ times, Beresnev and Trushin 1976) strengthening of material under a high confining pressure (it is this effect that makes possible the railgun launch of plastic bodies ~1 cm in size and ~1 g in mass to $V > 7$ km/s in a channel only $\approx 60$ cm long or hyperimpact ejection of asteroid-size fragments not only from the Moon or Mars, but from the Earth as well (Drobyshevski 1995 and refs. therein; for some details see also Sec. 8.2 below)). An excessive (confining) velocity head acts onto surface of a quasi-spherical body up to about its equator. Internal stresses in the trailing part of the body depend on a value of its deceleration. *Third*, the statistical pattern of the breakup and ablation processes should allow the existence of a certain amount of fragments that would reach the ground intact. Nobody has thus far made such an analysis accounting for the statistical character of the processes (although Svetsov (1996, 1998) being thinking in 1996 that all the fragments leaving the cloud would be evaporated by radiation of this cloud, in 1998 pointed out that some fragments of ~10 cm size could survive and fall at 5-10 km distance from epicenter, i.e. in a zone of the TP tree-fall). *Fourth*, the velocity head acts in total only on the front layer of the fragment ensemble, i.e., it is these fragments that will primarily be disintegrated later on. The velocity head of the flow acting on the next layer of larger fragments should, by definition, be lower. Here an ablation works mainly and whether it has a time for evaporating a rather great fragment needs to be studied. It appears that nobody carried out such a detailed consideration. So it can readily be seen that estimates are performed in "the most-favored-analysis" treatment for the EF breakup model accepted.

*Finally*, assuming the above dependence $V_t(V, \rho_a, \rho)$, an ensemble of fragments would meet a mass of air equal to its own mass $M$ in a path $L \sim (3M\rho/\pi\rho_a^2)^{1/3}$ (i.e., independent of the velocity), which for $M \approx 10^{12}$ g and $H \approx 8$ km amounts to about 20 km. But in order to slow down efficiently, it would have to travel a several times longer path (or to have a smaller $M$ and, correspondingly, energy $W$, - see below).

Gas-dynamic calculations were performed for such an EF by more than one researcher, and not only two- (Svetsov *et al* 1995) but three-dimensional as well, some of them with inclusion of radiation transport in air (Shurshalov 1982) and the onset of floating up of overheated plasma and/or ballistic ejection of an air plume containing the meteoroid mass to an altitude of $\geq 100$ km along the trajectory channel (Boslough and Crawford 1997, 2008a,b). One encounters here, however, certain problems in trying to reconcile results of these certainly high-level calculations with other considerations.

To begin with, attempts at substantiating an EF of a high enough energy ($W > 10$ Mt TNT) following from acoustic and seismic TP data fail as a rule. Therefore, the above researchers had to question the upper estimates ($W = 20 \div 50$ Mt TNT) and restrict themselves to $W \sim 10$ Mt TNT or even $\leq 5$ Mt TNT (Boslough and Crawford 2008a,b) and chose fairly steep trajectories with $\delta = 45°$ (Chyba *et al* 1993), $\delta = 40°$ (Rudenko and Utyuzhnikov 1999, Utyuzhnikov and Rudenko 2008) and $\delta = 35°$ (Boslough and Crawford 1997) to the horizon, which facilitates release of a higher energy in the denser low atmospheric layers (Shuvalov 2008). In actual fact, the angle of the trajectory could have hardly exceeded 20° (see below). Second, the tree-fall area turns out, as a rough estimate, an order of magnitude smaller (~300 km$^2$ by Rudenko and Utyuzhnikov 1999, Utyuzhnikov and Rudenko 2008) than actually found. On the other hand, Boslough and Crawford (1997) arrived at the desired pattern and area of the devastation for $\delta = 35^o$ under the assumption of the lowest possible strength of tree trunks using tree strength measurements performed by Florenskiy (1963) in 1961. Florenskiy quotes also fairly contradictory opinions of two experts, with one of them referring to a fire in the early XIX century after which robust 70÷100-y trees should have grown, whereas the other mentions a fire of 1889; this makes it unclear which of the opinions should be considered more trustworthy.

Furthermore, the tree-fall itself within the butterfly pattern area is highly nonuniform, thus prompting even a scenario of several consecutive air shocks (Goldine 1998, Vasilyev 2004, Chap. 2.2.4). One can not exclude also an existence of several tree-falls detached for hundred km from each other (Konenkin 1967). That seemingly contradicts to the EF approach if one does not imagine a common flight of several large meteoroids. On the other hand, such a tree-fall non-uniformity should indeed be expected, if for nothing else than the Taylor instability of the front of the meteoroid fragment cloud (e.g. Crawford et al 1995, Svetsov et al 1995) (see also below, Sec. 6.3).

### 4. Trajectory of the Tunguska bolide and its turn

Let us address now the specific features of the trajectory. Digressing for the moment from the SSW-NNE trajectory of Voznesenskii-Astapovich (with an azimuth $\varphi \approx 164^o$-$206^o$) deduced apparently from sound effects primarily (Astapovich 1933, 1951)[1], we arrive for the azimuth of the true trajectory as derived from eyewitness accounts as being apparently within $\varphi \approx 120^o \div 137^o$ (Krinov 1949, 1953, Bronshten 2000, Epiktetova 2008). Its inclination, according to the summary of Bronshten (2000), Vasilyev (2004) and Sekanina (1998), was $\delta \approx 0 \div 20^o$, whereas in EF calculations it was assumed equal to $35^o$ and even as large as $45^o$ (see above).

There are two extremely important and intriguing observations: (*i*) eyewitnesses insist that the bolide flew (Vasilyev 2004, Chap. 3.2.1) not exactly where the explosion occurred but rather slightly to the north of this point, and (*ii*) the symmetry axis of the butterfly tree-fall pattern around the epicenter of the explosion has, according to Fast (1967), an azimuth of $\varphi = 115^o$ and even $\varphi \approx 99^o$ (Fast *et al*. 1976) (an azimuth of $\approx 95^o$ is obtained also from the extended zone of scorched vegetation by Vorobyev and Demin (1976)). This prompts a suspicion that the trajectory underwent a turn. This opinion, however, is viewed, as a rule, with some irony and rejected as there having been no physical reasons for a turn. Gas dynamic interaction of such a massive meteoroid (even if fully fragmented) with rarefied air (even in the presence of a wind) would not be capable of turning noticeably the trajectory (see also below). Therefore, this observation, be it even only on a subconscious level, could not but confer only a slightly larger weight to the "eastern" initial trajectories (Bronshten 2000). Nevertheless, some researchers (we would call them scientifically unbiased, if this definition is applicable at all to enthusiasts), basing on objective available data (Epiktetova 2008), tend to believe that a turn of the original trajectory with $\varphi = 126^o$ by $\Delta\varphi \approx 10^o$ did indeed occur at a distance from the epicenter of somewhere about ~250÷300 km.

Assigning a certain aerodynamic lift-drag ratio (and a fairly large one, up to -4÷+2, from the standpoint of gas dynamics; for a sphere it is 0) to a body of density 0.01÷0.05 g/cm$^3$, as this is done by Korobeinikov *et al*. (1984), to obtain a downward deflection of the body at the end of the trajectory to substantiate $\delta \approx 35^o \div 45^o$, can hardly be justified within the physical models considered up to now (outside the artificial nature of the body). This characteristic can have only a strong enough body of a regular and stable shape.

---

[1] Interestingly, the more emotional perception of phenomena (primarily of sounds) by the western eyewitnesses (on the Angara river) compared with those located to the east (upstream of the Lower Tunguska river) (see Vasilyev 2004, Sec. 3.2.1) can be assigned to the well known waveguide acoustic effects in the stably stratified morning atmosphere; indeed, the difference in local time between these regions is 30'÷50', a noticeable headway for morning-time warming of the ground-level air mass in the east. A non-standard idea was discussed by Zabotin and Medvedev (2007). They believe the Tunguska meteoroid, having been captured by the Earth as due to its atmosphere drag, made several (two, most probably) orbital revolutions around the Earth before the explosion. The authors believe that could explain the TP different appreciations by the eastern and western witnesses, as well as the observed trajectory 'turn'.

## 5. On the nature of the SP comets

Until the mid-XX century, a period ear-marked by the beginning of nuclear bomb tests in the atmosphere, it was difficult to estimate the seismic, acoustic, magnetospheric etc. consequences of the explosion of the Tunguska meteoroid in the atmosphere. The only thing possible was to carry out comparisons with volcanic eruptions, for which quantitative estimates, in their turn, were also not reliable enough. And only combining the results obtained in nuclear tests with the concomitant development of methods of hypersonic gas dynamics could clarify to some extent a number of aspects in the physics of the TP, including its energetics, which illustrates favorably the extent to which coupling between close areas of science may become fruitful.

As for the numerous contradictory and as yet unclear manifestations of the TP, it appears worthwhile to quote here the opinion of Vasilyev *et al* (1976) that "we have yet to wait for development of a concept capable of satisfactorily accounting for the totality of available observations. This may imply that the conditions conducive to its appearance are not yet ripe enough"…

Indeed, we may note that each generation of researchers has been putting a different meaning even into the words of the cometary origin of the TP (Bronshten 2000, Vasilyev 2004). While Whipple (1934) and Astapovich (1951) considered the cometary nucleus to be a conglomerate of ice-covered stones, possibly even isolated from one another, this concept changes starting from 1950 to an icy nucleus with inclusions of meteoroid material (Whipple 1950). The problem of its consistency (a lump of snow, rubble pile, monolithic ice etc.) and density has not thus far reached a universally accepted and final solution (the density is assumed to be $\rho \sim 0.3 \div 1$ g/cm$^3$, although the overall trend seems to be accepting $\rho \rightarrow 1$ g/cm$^3$). The recent active cometary missions *Deep Impact* (*DI*) and *Stardust* (*SD*) do not appear to corroborate the idea of a comet as a condensate in the cold periphery of the Solar System (see Drobyshevski 2008a, and refs. therein).

We are going to show here that the new approach to understanding the origin of comets and of their manifestations which we have been developing for nearly three decades now (Drobyshevski 1980, 2008a) may provide a significant step toward revealing the nature of the TP. In contrast to the traditional condensation-sublimation concepts which, as is becoming presently increasingly more evident, are not substantiated by many reported observations, including the recent *DI* and *SD* missions, our approach assumes a planetary origin of cometary nuclei. This New Explosive Cosmogony (NEC) of comets and of a number of other minor bodies of the Solar System (asteroids, the Troyans, irregular satellites, satellites of Mars, Saturn's rings etc.) believes them to be ejected in global explosions (possibly only seven or eight during the existence of the Solar System) of very thick (up to ~800 km) icy envelopes of Ganymede- and Titan-type bodies (for more details and refs. see Drobyshevski 1980, 1986, 1989, 1997, 2000, 2008a). The ices consisting of H$_2$O, primitive organics etc., with numerous mineral inclusions up to a few meters in size, are saturated *in the form of a solid solution* by products of their bulk electrolysis, 2H$_2$+O$_2$, to 15÷20 wt.%, which makes them capable of detonation.

The nuclei of SP comets, which are actually unexploded fragments of the outermost layers of the envelope, also contain 2H$_2$+O$_2$ in a close to critical concentration, which accounts for practically the whole totality of the cometary manifestations whose origin had remained for a long time mysterious (e.g., bursts and breakups of the nuclei which correlate with solar activity, the appearance of ions and radicals in the close vicinity of the nucleus, etc.) (Drobyshevski 1988b). It presently appears that it is these new concepts implying a possibility of a chemical explosion of ices of a cometary nucleus in the Earth's atmosphere that provide the heretofore lacking link capable of solving the TP problems.

## 6. The TP scenario with chemical detonation of $2H_2+O_2$-containing cometary ices

Indeed, a close-to-tangent entry into the Earth's atmosphere of a small icy comet nucleus (with $\rho \approx 1$ g/cm$^3$), which is possibly already strongly surface-deactivated (dormant) and coated by a fairly loose heat-insulating layer of "sand" grains bonded together by polymerized organics, would bring about, first of all, loss of this coating (an analog of the beginning of the EF), thus increasing the bolide size to 0.5÷2 km in diameter (Astapovich 1934) (the diameter of conventional bolides may be as large as a few hundred meters). This can be followed by the onset of the EF itself, i.e., destruction of the ice by the velocity head. The resultant compression, the onset of breakup etc., including penetration of the impact-heated air over cracks inside the (icy) body, would inject an additional energy into its bulk, with initiation of the subsequent detonation of the $2H_2+O_2$ + organics mixture dissolved in ice, possibly, independently in several different pockets and/or veins. Significantly, the energy of the chemical explosion is substantially lower than the kinetic energy of the body. By dispersing the body, the explosion provides conditions favorable for a fast gas-dynamic transformation of the energy of its motion into the energy of the eventual volume of overheated air mixed with the detonation products.

*6.1. Detonation of a part of the body as the cause of the turn of the final trajectory*

The totality of observations suggest that the various manifestations of the TP originate from detonation of only a part of the fairly large (possibly, many hundreds of meters in size) true cometary nucleus.

As follows from the images of the nuclei of SP comets obtained in the two recent decades, they have, on the whole, a fairly irregular, fragmentary shape and a layered structure (see refs. in Drobyshevski 2008a). The layers, rather than being concentric, are planar and longitudinal; approximating the shape of a nucleus with an ellipsoid, they are normal to its minor axis as a rule. This structure is revealed in Phobos, the Martian satellite, which within the NEC is considered, similar to Deimos, to be a fragment produced in the explosion of a larger, ~200-km icy fragment and captured into a near-Martian orbit (Drobyshevski 1988a). The layered structure is a consequence of the geological processes (solid-state convection) that occurred in the icy mantle of a Ganymede-type parent body. On Phobos, this structure, which is normal to the minor axis of its ellipsoid and parallel to the major axis, is manifested by group C grooves. It appears only natural to assume that the $2H_2+O_2$ concentration, just as the strength characteristics of the material, also follow to some extent this layered structure. It is likewise clear that the breakup of the body under the action of velocity head, just as the detonation of ice containing $2H_2+O_2$, would likewise start in the layer oriented approximately along the flight direction, more specifically, in the part of the layer close to the leading part of the body which is heated by hot air slowed down by the bow shock wave.

The detonation wave propagates through the $2H_2+O_2$-containing ice with a velocity $V_{det} \approx$ 5.2÷5.5 km/s, at a pressure behind it of $p_{det} \approx (5.7 \div 6.5) \times 10^{10}$ dyne/cm$^2$ (Drobyshevski 1986). If the plane surface layer of a typical explosive with a mass $m$ detonates completely, it will acquire in repulsion from a much more massive body a momentum $I = mV_r \approx 8mV_{det}/27$ (Baum *et al* 1975). In our particular case, the material of the exploded layer (with an area, say, of ~200×200 m$^2$ and thickness of 20 m) will depart from the substantially more massive unexploded part ~200÷500 m in size with a mean-mass recoil velocity $V_r \approx 1.54 \div 1.63$ km/s. For a total velocity $V = 20$ km/s, this corresponds to a turn of the center-of- mass trajectory of the rapidly expanding products of detonation through $\Delta\varphi \approx 4.5^o$. For $V = 11.2$ km/s, the turn would be $\Delta\varphi \approx 8.0^o$. To corroborate the data of Epiktetova (2008), we would have to obtain $\Delta\varphi \approx 126^o$-$115^o = 11^o$, but, in view of the fact that the error of determination of the angle of turn basing on eyewitness accounts may reach a few degrees, say, even ±12$^o$ by Zotkin and Tshigorin (1988), one may content oneself with the above result.

*6.1.1. Possible effect of the lift-drag ratio of the detonated layer on the additional turn of its trajectory*

It appears appropriate to recall here the abovementioned results of Korobeinikov *et al* (1984) (which may appear at first glance to be not very physical) about a possible role of the lift-drag ratio in the evolution of the Tunguska meteoroid trajectory. The calculations of these authors acquire a certain physical meaning if one takes into account that detonation of a near-surface layer oriented along the direction of motion starts from the front part of the meteoroid. Due to the explosion, the plane layer of hot detonation products ($H_2O$ vapors mainly) which is still dense is first repulsed by the pressure of the explosion, to gradually begin, *now by its inertia*, to drift away from the main body, to reveal an ever growing gap between this layer and the body, which is oriented in the direction of motion of the original body and keeps being a somewhat evacuated (we may recall here the vacuum created inside the expanding sphere of detonation products in the case of a spherical explosion (see Baum *et al* 1975) accompanied in this particular situation by condensation of the adiabatically expanding water vapors). It is possible that the air incident on the body enters this kind of an air intake, is slowed down and drives the detonated layer away from the unexploded part by imparting an additional transverse momentum to the layer. Until it is not destroyed, the layer acquires, as it were, an aerodynamic force, which is particularly high as long as it remains close to the main body ("wing-in-ground effect" - the effect of enhancement of the lifting force of a wing if it moves close to the ground is well known in aerodynamics and is employed in development, say, of ecranoplanes). If the detonated layer of material at least doubles its $V_r$ through these processes (which is possible if the layer retains its integrity and continues to accelerate through the action of impact air pressure over a distance of 10÷20 its thicknesses), this will add up in its effect to a ~$10^o$ transverse deflection of the mean-mass trajectory of the expanding detonation products at 250÷300 km from the epicenter.

*6.2. On the consequences of the turn the trajectory of a part of the cometary nucleus undergoes as a result of detonation*

This suggests that somewhere about 200÷250 km north-west of the "Kulik epicenter" there could exist one more epicenter, possibly, an impact crater produced by the fall of a larger one, 200÷500 m in size, unexploded (or not so efficiently braked in the EF) fragment. The absence of any other relevant manifestations makes, however, this possibility, just as the possibility of proving two- more tree-falls to be associated with the TP, hardly acceptable (see, however, Sec. 9 below).[2]

One could recall here reports of several (as a rule, three to four) strong sonic shocks following one another with intervals of tens of seconds. These shocks could be produced by the electrophonic effect, the shock wave generated by the hypersonic bolide flight along its trajectory, the impact caused by a high-altitude NEC explosion of the near-surface planar layer of the original icy body that has changed the azimuth of the trajectory of its detonation products by ~$5^o$÷$10^o$, the sound of the fall of the products of this detonation together with the concomitant mass of the trapped overheated air on the surface, a surface explosion with the Kulik tree-fall and, finally, a possible fall of the large remnant body somewhere to the north-west of the devastation. One cannot exclude also multiple reflections of sound from clouds, the waveguide atmospheric effects and so on.

---

[2] We become confronted here by the problem not only of the azimuth but of the true inclination of the resultant trajectories, as well as of partitioning of the energies (seismic, acoustic and others) between these falls and so on. One cannot exclude the possibility that the fall of the unexploded fragment (or even of several of them) which occurred to the north or north-west of the Kulik tree-fall and is rather closer to the Angara river than to the upper reaches of the Lower Tunguska that adjoin the Lena river (Mironovo village) also initiated the abovementioned excited accounts of the Angara eyewitnesses (and, possibly, the SSW-NNE trajectory of Voznesenskii-Astapovich).

Hopes for obtaining here immediately from a high-altitude explosion of the layer also a ~10° increase of the trajectory inclination, from, say, $\delta \approx 20°$ to $\delta \approx 30°$ as an extra bonus appear groundless for purely geometric considerations; indeed, from an altitude of 10 km cited usually for the EF scenario (see above) and a trajectory inclination of 10°, the body would fall on the ground already on passing a distance of 57 km (to strike the Earth in 300 km, the body with $\delta \approx 10°$ would have to be at an altitude of 50 km, and for $\delta = 30°$, at $H = 173$ km; using the relation from Zotkin and Tshigorin (1988), Epiktetova (2008) arrived at $\delta = 28°\div29°$). Some discrepancies are obvious, although one could naturally try to find also a combination of parameters that would suit both scenarios. On the other hand, what follows from this is rather that *one should look for other ways* of reconciling the small inclination of the trajectory with the forest fall parameters etc., an approach we are going to develop basing again on the NEC.

### 6.3. Correlation of some EF implications with the NEC inferences: the butterfly pattern and softening of the constraints on the TP energetics

Any physical ideas concerning the nature of the TP reach the critical stage when their implications have to be correlated with the "witness plate" data, i.e., with the butterfly pattern of tree-fall over an area of 2150 km$^2$. It is worthwhile to note that this butterfly has no head.

As already mentioned (see Sec. 3.1), the butterfly pattern can be modeled numerically basing on the EF concept, but only just, i.e., if one assumes that (*i*) the trees in 1908 had the lowest possible strength - otherwise the fall area becomes an order of magnitude smaller, and (*ii*) the inclination of the meteoroid trajectory to a horizontal plane $\delta$ is fairly large - 35°, 40°, or 45°, whereas direct observations suggested $\delta = 0\div20°$ (Bronshten 2000, Vasilyev 2004, Sekanina 1998) and, the largest cited, $\delta = 20°\pm6°$ (Zotkin and Tshigorin 1988).[3] We should like to stress once again that accepting such large angles forces the authors to limit the TP energy in their consideration by the level $W \leq 15$ Mt TNT.

To clarify the situation, compare the scenarios of meteoroid energy transfer to air in two cases, namely, the EF and a chemical explosion of an icy body saturated with $2H_2+O_2$. The differences consist actually in that the radically different initial conditions lead one to different final spatial distributions of the material of the dispersed meteoroid and of the energy contained in this material.

In *the first case*, one assumes that the velocity head $\rho_a V^2/2 = 10^9$ dyne/cm$^2$ (for $V = 20$ km/s and air density $\rho_a = 0.5\times10^{-3}$ g/cm$^3$ at an altitude $H = 8$ km), a figure in excess of the material strength (is this indeed so or not, is another question, see above), breaks up the body into fragments, so that their ensemble begins to behave as a liquid. Initially it spreads out laterally, perpendicular to the direction of motion, with a velocity $V_t \approx V(\rho_a/\rho)^{1/2} \approx 450$ m/s (for $\rho = 1$ g/cm$^3$). Here the passive entertainment of fragments and their ablation products by air has place. To interact with the mass of air equal to that of the body ($M \sim 10^{12}$ g, $W = 50$ Mt TNT at $V = 20$ km/s), such a *plane ensemble* of fragments, as we have seen, has to travel a distance $L \sim 20$ km (for $M = 10^{11}$ g, $L \sim 10$ km). The path ensuring effective slowing down and energy transfer to air should be about twice this distance, yielding ~2 s for the slowing-down time. In this time, the diameter of the disk reaches about 2 km, and at the center it thins down to become eventually a ring-shaped vortex (Boslough and Crawford 2008b) or a series of such vortices.

Note that in the case of a smoothly inclined trajectory ($\delta < 20°$) the upper part of the ensemble of the EF fragments moves at a lower $V_t$ than the lower one because of $\rho_a$ decreasing with height. Accordingly, the density (and the longitudinal velocity) of the ensemble will be here somewhat larger than below. The drop of this part on the ground would add the head to the butterfly, which is in conflict with observations; this is why the calculations based on the EF concept have to be restricted to steep trajectories and, as a consequence, to lower values of $W$. On the other hand, for the altitude at which the trajectory turns, $H = 50$ km, we have $\rho_a = 10^{-6}$

---

[3] The error value ±6° for the inclination $\delta$ instead of ±12° given by Zotkin and Tshigorin (1988) follows from their notion that the trajectory azimuth has twice greater error than its inclination; it was approved by Zotkin after our discussion of the issue in December, 2008.

g/cm$^3$ (for $H$ = 150 km, $\rho_a$ = 2×10$^{-12}$ g/cm$^3$), which reduces the velocity head to such a small level as to make the EF concept totally unacceptable.

We arrive at *a different scenario and results* if the body explodes at an altitude $H$ (we again assume $H$ = 5÷10 km for estimates) in an initially nearly spherically symmetric pattern, be it even with an energy of explosion lower by far than its kinetic energy (the consequences of a chemical explosion of a material of an already evaporated cometary nucleus mixed with air were discussed by Tsymbal and Schnitke 1988, and Kondratyev *et al* 1988).

In the case of an exploding icy body which is of interest to us here, the detonation wave velocity $V_{det}$ ≈ 5.2÷5.5 km/s (Drobyshevski 1986), so that the velocity of transverse expansion of the detonation products $V_t$ ~ 2 km/s is prescribed and, thus, does not depend on $V$ in an explicit way, as in the EF case, although it drops to about one half this value at a distance of ~500÷800 m because of the resistance of the air with $\rho_a$ ≈ 0.5×10$^{-3}$ g/cm$^3$ ($H$ = 8 km). Moreover, there are grounds to believe that here $V_t$ should be the larger, the smaller is $\rho_a$ (in the limit of $\rho_a$ = 0, the velocity of the front of the explosion products, water vapor with the initial temperature $T$ = 900÷1000 K, expanding into vacuum will be as high as ~5 km/s, exactly what was observed in the *DI* experiment (Drobyshevski *et al* 2007)). Therefore, the exploded body will meet at an altitude of 7 km, for $V_t$ ~ 2 km/s and $V$ = 20 km/s, an air mass equal to its mass $M$ (≈ 10$^{12}$ g) along the path $L$ ~ $(3MV^2/\pi\rho_a V_t^2)^{1/3}$ ≈ 5 km, i.e., it will transfer all of its kinetic energy (including the high $W$ ~ 50 Mt TNT) in a few fractions of a second (e.g., ~0.5 s, not in some seconds typical for the EF). Unlike the standard nuclear explosion conditions, all of this mass, continuing to expand and radiate, will move with a velocity of ~10 km/s.

We stress one more that the altitude of 5÷10 km at which $\rho_a$ ≈ 0.5×10$^{-3}$ g/cm$^3$ is not very critical. We do not have to destroy the body by the velocity head; indeed, electrolyzed cometary ices are capable of exploding at any altitude, say, at $H$ ~ 50 km.

Practically all of the exploded mass is confined initially to a fairly thin, nearly *spherical layer* expanding with $V_t$ (starting with distances $r$ noticeably in excess of the original size of the body, say, for $r$ > 1 km, one may, for the sake of simplicity, invoke the point, spherical explosion approximation). The oncoming hypersonic air flow pushes inward the front of this sphere while the "side walls", which are shaped as a short "cylinder" of a length roughly equal to its diameter and have a larger mass per unit cross section (with respect to the oncoming flow), continue their motion. For a smoothly sloping trajectory with $\delta$ ≤ arc tan($V_t/V$) ≈ 5°÷10°, the upper part of the cylinder, in contrast to the EF case, shots up into more rarefied strata of the atmosphere overtaking the slower-moving side walls, and all the more so, the bottom of the cylinder. The only thing left of the cylinder is actually a kind of a trough, which gradually expands, in this stage under the action of the pressure of the gas flowing through it with a temperature higher than that of the outside air. It is the gas dynamic interaction of the trough with the ground that results in the forest fall shaped as a headless butterfly. The distance between the zones of maximum shock pressure at its wings is ~30 km, which corresponds to the average velocity of the trough side wall expansion 2$V_t$ ≈ 0.1·$V$ (≈2 km/s for $V$ ≈ 20 km/s)[4], if this expansion started exactly at the time of the trajectory turn. Significantly, the fall on the ground of the side walls of the trough should give rise to two columns of fire, exactly what eyewitness Romanov from Nizhnee Ilimskoe first saw, and only thereafter he heard two shocks (Krinov 1949).

Interestingly, the vertical component of the velocity of the body, ~1 km/s at $\delta$ ≈ 5°÷10°, is comparable with the average value of $V_t$. This means that as the trough moves down, its bottom will slightly push up and lag behind its side walls (a process similar in some respect to behavior of the front of the initial spherical layer of the detonation products, - see the beginning of the previous paragraph). Therefore, the front shock wave and the material itself of the trough bottom will reach the ground a bit later than those of the side walls. Combined with the later arrival of the stronger slowed-down front and trailing lids of the cylinder, which are by no means plane, if only for the development in them of Taylor instabilities, all this is capable of producing the pattern of tree-fall from several shocks (impacts), as described above in Sec. 3.1 (its last paragraph).

---

[4] Note that the EF would give a twice smaller $V_t$ value; that is why the EF approach faces, e.g., problems with obtaining the observed tree-fall area.

### 7. A note on the TP and on the SL-9 impact onto Jupiter

It appears of interest to estimate the extent to which the reasoning developed above could be applied to the fall of fragments of SL-9 on Jupiter, particularly if one takes into account the differences in velocity (and energetics) of the fall (~20÷30 km/s for the Tunguska meteoroid and ~60 km/s for SL-9). The velocity of the initial transverse expansion of the ensemble of fragments produced in EF for TP for $\rho \approx 1$ g/cm$^3$ at $H \approx 8$ km is, as we have seen, $V_t \approx 450$ m/s, which is a few times lower than that in the case of chemical detonation. For Jupiter at a 1-bar level, $V_t \approx V(\rho_a/\rho)^{1/2} \approx 600$ m/s, i.e., higher by one third only. It would thus seem that one should have to include into consideration the more efficient interaction with Jupiter's atmosphere of the material of the SL-9 fragments initiated by the explosion of their electrolyzed ices. On the other hand, judging from their very weak cometary manifestations, the SL-9 fragments lost their ices earlier, possibly in the breakup of the primary nucleus (see discussion in Drobyshevski 1997). In this case, the concepts of Klumov *et al* (1994) and Crawford *et al* (1995) concerning the interaction of SL-9 with Jupiter's atmosphere do not require, unlike the TP, a sizable revision.

### 8. Possible comet impact cause of the late Pleistocene climatic catastrophe

In the recent decade, evidence has been building up that the Young Dryas global cooling that had started ~12.9 ka ago and had brought megafaunal extinction was actually initiated by a fall and/or explosion of a cometary nucleus somewhere in the Northern hemisphere, probably, over North America (Firestone *et al* 2007, Kennett *et al* 2008). This assumption is corroborated by the thin layer of sediments, which contains, similar to the peat moors near the epicenter of the Tunguska fall in 1908, indications of the presence of extraterrestrial (cometary?) material that had been subjected to high pressures and temperatures (for instance, nanodiamonds (Kennett *et al* 2009)). In view of the fact that this layer can be unearthed all over the North America, can be identified with similar findings in Europe (e.g., Lommel in Belgium), and reveals aftereffects of large-scale (possibly, continent-wide) fires, this event was much more powerful than the TP and was accompanied by ejection into the atmosphere of great amounts of dust and soot.

The corresponding impact crater has not, however, been located in the North America thus far. The absence of the crater could also be treated as an evidence for the fall of a meteoroid (a comet nucleus) in two ways. On the one hand, it can be assumed that the amount of dust loaded into the atmosphere by the cometary nucleus itself from outside was large enough to initiate a climatic catastrophe. Straightforward estimates reveal, however, that such a large nucleus reaching the surface would produce a crater whose ejecta would exceed by far the mass of the nucleus itself. One has therefore to understand where this impact crater we are interested in should or could be, and what it should be like.

#### 8.1. Plate-like and two-bowl comet-impact craters

The diameter of an impact crater $D$ in typical Earth's rock surface is estimated as $D = KW^{1/3.4}$ cm (where K = 0.016, $W$ is in ergs, e.g., Shoemaker and Wolfe (1982)), and its transient depth $h_0 \approx D/4$. It is assumed that the impactor size $d_0 \ll D$. Then the resulted crater's bowl-type shape is similar to the one created by an outburst of undersurface-located explosive charge.

In the case of explosion of the $2H_2+O_2$ solid solution in cometary ices at altitude $H$, a situation may occur when the products of detonation of the body (or of the breakup of the meteoroid by the velocity head, i.e., in the EF; here $V_t \leq 1$ km/s), in expanding with $V_t$ ($\approx 2$ km/s), will reach the surface (irrespective of how efficiently they are braked down by air resistance) in a cloud with a transverse size larger than $D$ as defined by the expression above.

Obviously enough, the collision would not give rise here to a clearly pronounced bowl-type crater; one can expect here only insignificant displacements of surface layers, exactly what was observed in the TP (Krinov 1949; a substantial part was played here of course by the air flow dragged along by the directed motion of the detonation products). A well pronounced crater with ejecta may form if the expanding detonation products will grow at the time they reach the Earth to the size $d = 2V_tH/V < D$ (to gain an overall idea of the scale and physics of the processes involved, we are neglecting in what follows the lateral slowing down and drag by the air and assume $V_t \approx$ const).

Whence it follows that a crater first appears as a fairly plane depression if the original size of the nucleus $d_0 \geq [(12/\pi\rho) \cdot (2H/K)^{3.4} \cdot V_t^{3.4} \cdot V^{-5.4}]^{1/3}$. For $H = 10$ km, $\rho = 1$ g/cm$^3$, $V_t = 2$ km/s and $V = 20$ km/s, one obtains $d_0 \geq 110$ m (mass of the nucleus $0.7 \times 10^{12}$ g, $W = 33$ Mt TNT). A fall of such a compact nucleus would produce a crater with $D = 2$ km, but the detonation-induced expansion of the nucleus material would spread it over the surface in a circle exactly 2 km in diameter, so that the crater should be pretty shallow and hardly discernible already a short time thereafter.

It follows immediately that (*i*) if the Tunguska body with $M \sim 10^{12}$ g had exploded at an altitude $H \sim 10$ km, it would not have created, even if one disregards the braking by air, a full-scale crater with the corresponding ejection of rocky material, and that (*ii*) even an exploded body with $M \sim 10^{13} \div 10^{14}$ g ($d_0 = 270$ m, $W = 500$ Mt TNT, and for $d_0 = 570$ m, $W = 5000$ Mt TNT) would have produced a shallow flat-bottom, plate-like crater; the atmosphere would have been loaded with a lot of dust from it, but the crater would hardly have been identifiable, if we use for reference, say, the bowl-like Arizona crater.

One can readily calculate that taking into account the dragged air mass would require an increase of the nucleus in size from $d_0 \approx 100$ m to $d_0 \approx 500$ m, i.e., about fivefold, in order that, having been exploded, it could reach the Earth with a larger part of its original energy left non-dissipated (in our particular case, it is $W \approx 4000$ Mt TNT). This increases $D$ to $\sim 6 \div 8$ km, but the crater would still be not so large as if a compact body had struck the ground. The ejected dust could now, however, trigger a climatic catastrophe $\sim 12.9$ ka ago, first at least in one (Northern or Southern) hemisphere (see also the next Sec. 8.2).

The above reasoning suggests that our present understanding of the consequences of crater-producing impacts on the Earth are plagued with a gap; indeed, even fairly large ($\sim 200 \div 500$ m and, possibly, even 1-km sized) comet bodies, in which the solid solution of $2H_2+O_2$ with organics in ice detonates at an altitude, strike the Earth and produce ejection of material into the atmosphere, with the ensuing climatic consequences, do not apparently create the familiar deep, bowl-shaped crater, because the impact momentum spreads over a large area.

And what is interesting finally, in the case of the partial comet nucleus explosion (similar to the Tunguska case under consideration), the two-bowl crater has to be created, viz., the deep central one caused by the unexploded part fall and the surrounding shallow plate-like crater resulted from the hypervelocity impact of products of the $2H_2+O_2$-containing another icy part detonation at the high altitude.

*8.2. The fire-sowing skies*

The widespread development of wildfires over the North America territory, along with an absence of an appropriate impact crater there, forced Firestone *et al* (2007) and Kennett *et al* (2008) to suppose that the continental-scale wildfires were caused by a parallel outfall of a swarm of comet nuclei and their atmospheric explosions (due to EFs, for example).

Nevertheless, the Earth surface appears to bear an impact crater responsible for the catastrophe under consideration. This is the Zhamanshin crater (48.4° N, 61.0° E) in Kazakhstan. It consists actually of a shallow crater with $D \approx 13$ km and depth $h \approx 200$ m enclosing a somewhat deeper ($h \approx 250\text{-}300$ m) $D \approx 5.5 \div 6$ km crater. Usually it is ascribed to have $\sim 1$ Ma age, being dated by accompanying impact glasses (tektites). It seems the small value $h/D \sim 0.02$, as being possibly due to erosion, is favoring such a great age also.

However, in 1980's Izokh (1990) concluded, basing on stratigraphic data, that the crater age is about 10 ka (by radiocarbon dating). At the same time he related its origin with the Late Pleistocene catastrophe. The rim bank of the crater is composed of loose, non-durable material. The radiation track age of basic glasses (viz., so called zhamanshinites, deliberately belonging to the crater) is <0.1 Ma (Izokh *et al* 1990). If it is so, then the small-value $h/D$ is not due to erosion but corresponds, just as the whole manifestation of the Zhamanshin crater, to the above discussed two-component flat craters created by an impactor partially exploded in atmosphere, viz. by the comet nucleus.

In the neighborhood of the Zhamanshin crater there are several scores of well-preserved (in Oligocene sands) smaller satellite craters $D \approx 50 \div 200$ m of the same age (Boyko and Sazhnov 1981). Most probably, they are produced by outfall of rocky fragments ejected from the main crater. Similar secondary craters are well known on the Moon and other planets. The ~10 ka age near-crater strata in Kazakhstan also demonstrate signs of wildfires.

It remains factually to understand how could the Zhamanshin event initiate the wildfires not in its immediate vicinity only, but in the opposite hemi-sphere too?

One can hardly imagine the atmospheric shock wave could reach another hemi-sphere to heat the air there up to a temperature of a wood ignition (>800 K), - this would result in shedding a significant part of the atmosphere itself.

It is much simpler to assume that a lot of rather large rock fragments were ejected gasdynamically from the main crater, primarily along the rarefied initial trajectory channel in atmosphere, in space. The maximum size for the ejected fragments' distribution is defined as $\varnothing_{max} \approx 2\sigma_{max}D/3\rho V_{esc}^2$ (Drobyshevski 1990, 1995).

As that was pointed out in 1990, such hyper-impact rocky ejecta from the Earth are able to explain not only origin of some "Martian" SNC meteorites and their properties, but a cause of observed excess of near-Earth asteroids (Drobyshevski 1990, 1995, 2002). For dunite $\sigma = 2.5 \times 10^9$ dyne/cm$^2$, for gabbro – $4.2 \times 10^9$ dyne/cm$^2$, for diabase, it is $8.3 \times 10^9$ dyne/cm$^2$. With taking into account an effect of the material strengthening under conditions of great confining pressures, these $\sigma$ values can be increased up to $\sigma_{max} \sim 10^{11}$ dyne/cm$^2$ (Beresnev and Trushin 1976). Then, even at $\sigma_{max} = 10^{10}$ dyne/cm$^2$, and $\rho = 3$ g/cm$^3$ one obtains for $D = 13$ km Zhamanshin crater the ejected fragments of $\varnothing_{max} \approx 25$ m for their $V = V_{esc} = 11.2$ km/s! The plate-like shape of this crater makes an ejection process from it less effective, - so that would lower $\varnothing_{max}$ down to ~10 m. Nevertheless, possibly, just this impact does explain the modern excess of ≤10 m size near-Earth asteroids (Drobyshevski 1990, 1995, and refs. therein).

The ejection of such and smaller (even dusty) fragments into space gave rise firstly to the multiplicity of "ballistic missiles", whose return hypervelocity passage through the Earth's atmosphere heated their surface up to 2000÷3000 K (by Kuznetsov (1965), the air temperature at $H = 0$ km behind the shock wave with $V_s = 11.2$ km/s is 17000 K). That made them to be highly efficient igniters of wide-spread local wildfires on different continents, i.e. not necessarily in North America where paleo-fires have received the most study (e.g. Marlon *et al* 2009, and refs. therein), but in other locations as well. Falls of fragments ejected in the near-Earth heliocentric orbits continued with some natural gradual decrease in their frequency even during millennia after the primary comet impact. This scenario seems to be able to soften arguments posed by Marlon *et al* (2009) against the Extra-Terrestrial cause of the Younger Dryas. They argue the wildfires were not continent-wide single event, but while being numerous, their cases were localized and extended in time, – just that one could suspect basing on our approach. Moreover, data by Marlon *et al* (2009) on the fire frequency in North America allow to assume that the comet impact (Zhamanshin?) as such took place ~13.2 ka ago. By dust-loading of atmosphere in the Northern hemisphere together with causing actually a sudden jump (not a breakpoint in distribution) in number of soot-producing wildfires (see Figs. 1B and 1C in Marlon *et al* 2009)[5], the impact could firstly result in short ~150-200 yr Intra-Ållerød Cold

---

[5] Significantly, a more appropriate approximation for the charcoal influx at $15000 \geq Age \geq 13000$ y BP presented in these Figs. would be not 6.6102 - 0.000512×*Age*, as specified in Supporting Information by Marlon *et al* (2009),

Period and further on in successive abrupt megafaunal extinction and advent of the Younger Dryas cooling. All said above urges a necessity of scrutinized study of the Zhamanshin impact event.[6]

## 9. Conclusion

We have thus seen that the NEC, which has already demonstrated its validity and potential in the interpretation and prediction of many properties of minor and related bodies of the Solar System, offers apparently a possibility of removing the last lacunas in our understanding of the TP; in other words, following Vasilyev *et al* (1976), "…this may imply that the conditions… are… ripe enough"…

This approach permits one to understand the reasons for strong seismic and acoustic phenomena in the absence of a large impact crater, while the fast ejection through the trajectory channel into the upper atmosphere of a plume of hot ionized air together with the material of the exploded (detonated) meteoroid (Boslough and Crawford 1997) is capable of accounting both for the origin of light nights within a band from Tashkent to St.-Petersburg wide and extending to the Atlantic Ocean, and for the geomagnetic perturbations. The latter resemble in their parameters, on the whole, the perturbations initiated by high-altitude nuclear explosions, the only difference being that the former had greater amplitude and lasted 2÷5 hours rather than ~1 h (Kovalevsky 1963, Vasilyev 2004, Chap. 2.2.3), a feature easy to interpret, because nuclear explosions inject into the atmosphere 3 to 4 orders of magnitude less of highly ionizable material (with an ionization potential <10 eV).

Besides the biological aspects, important problems of the nature of silicate and magnetite spherules and possible chemical and isotopic shifts, there are only three unclear physical aspects in the TP that still remained thus far unresolved:

(1) The very high efficiency of transfer of the meteoroid kinetic energy and momentum to the air over a short distance and in such a short time that its expansion caused by injection of this energy produced a blast comparable with nuclear explosions;

(2) The need of inclusion into consideration within the EF concept of too steep trajectories ($\delta \geq 35^\circ$) in order to model the observed pattern and area of forest fall (assuming also the low strength of the trees) and scorching of the trees by radiation;

(3) The turn of the trajectory by $\Delta\varphi \sim 10^\circ$ to the west in the horizontal plane high in the atmosphere ($H \geq 10$ km, possibly, even $H > 50$ km).

It has been shown that an analysis of the cometary origin of the TP, now assisted by a new insight into the nature of the comets themselves, is apparently capable of answering these last questions.

This suggests, however, several more implications.

(1) It is possible that some other parts of the cometary nucleus (for instance, pockets with excess $2H_2+O_2$ concentration) also did explode, but their material dropped out a bit earlier and closer to the direction of the initial trajectory. It is these parts that could produce two more tree-falls to the north of the latter, at a distance of 90 and 120 km to the east of the Kulik's devastation area. In discussing their origin, Konenkin (1967) reported their trajectory azimuth as $\varphi \approx 120^\circ$, without taking into account its final turn. If one includes the turn, both falls will lie practically on the continuation of the initial trajectory. Konenkin did not succeed in identifying the seismic events associated with these tree-falls, because the relevant seismic waves, if they

---

but rather 5.6942 - 0.0004575×*Age*, i.e., the regression describing the influx jump at the transition to the Younger Dryas.

[6] It is presently believed that the K-T transition was initiated by a fall that produced the Chicxulub crater ($D$ = 180 km) ~65 Ma ago. The subsequent falls of asteroid-size fragments up to ~1-2 km in size ejected by this impact into near-Earth orbits certainly could have produced the Boltysh ($D$ = 24 km) and Silverpit ($D$ = 3 km) craters, while the prolonged fall of numerous smaller fragments initiated local fires on different continents; this could account for the strange differences in combustion products in various places, which were pointed, for example, by Belcher *et al* (2009).

had been produced, should have reached Irkutsk and other seismic stations slightly after the arrival of the wave front from the Kulik event, i.e., within their wave packet, thus making it hardly possible to disentangle the overall superposition of fairly weak signals.

(2) The ices of cometary nuclei may contain, besides fine sand, rocks up to ~5 m in size (Drobyshevski 1980). The drop of such a rock could easily account for the surface crater 8 km north-west of the Kulik tree-fall (lake Cheko, Longo 2007).

Thus, the NEC is seen to be capable of shedding light on a number of finer points in the TP. Remarkably, it offers a possibility of explaining and combining in its framework the observations and ideas that at first glance would appear incompatible and hardly probable, such as (*i*) the "explosion in flight" concept, including early appreciation of the need for a very low density of the primary body (Petrov and Stulov 1975), and (*ii*) the idea of a chemical explosion of the object (in addition to our NEC, see Tsynbal and Schnitke 1988, Kondratyev *et al* 1988), (*iii*) a turn of the trajectory (with a possible consideration of the lift-drag ratio for the detonated part of the body that could play an essential role while it lasted) and (*iv*), possibly, the impact origin of lake Cheko (Longo 2007).

Some minor aspects of the TP need certainly a thorough analysis within the frame of the NEC. Particularly, presence of different type nanodiamonds in peat of Tunguska swamps (Kvasnitsa *et al* 1979; Sobotovich *et al* 1983) and in the Younger Dryas strata (Kennett *et al* 2009) could hopefully shed light onto thermodynamic conditions of their appearance, say, on presence of pressures characteristic of detonation, not of the air velocity head ($p_{det} \approx 6\times10^{10}$ dyne/cm$^2$ >> $\rho_a V^2/2 \approx 10^9$ dyne/cm$^2$) or of the solid body impacts only. We are confident that the NEC would only benefit from such an analysis.

One more essential implication of the above discussion.

The fall on the Earth of the main, unexploded part of the comet P/Tunguska 1908 nucleus, 200÷500 m in size, up to ~5×10$^{13}$ g in mass and with an energy $W \approx 250\div3000$ Mt TNT would have resulted in creation of a crater with a diameter of ~3.5÷8 km. While a nuclear winter would not have set in on the Earth (this would have required a fall with $W \sim 10^5\div10^6$ Mt TNT (Drobyshevski 1989, and refs. therein)), the immense clouds of dust injected into the atmosphere and outfall**s** of the fire-igniting up to ~10 m size secondary rocky meteoroids would be certain to produce a catastrophic impact on the climate of the Northern hemisphere, with an ensuing global cooling similar to Younger Dryas and food and economic crisis. Mankind had been lucky indeed that this body traversed the Earth's atmosphere at an altitude of 5÷10 km and wandered back into space in 1908. People had actually witnessed realization of one of the scenarios of Vernadskiy (1932)[7]. We have got used to saying (Astapovich 1934) that if the TP had happened 4.7 hours later (it would seem to be such a trifle!), St.-Petersburg, which is located at the same latitude but 4000 km to the west, would have perished, and history of the XX century would have taken a different turn. But for the Tunguska event of 1908 to have become a crisis on the global scale, it would suffice that the trajectory of the cometary nucleus at the distance of the Moon differed in direction from the real course of the body by as little as ~5 seconds of arc! This is a good warning and a timely suggestion that Man be more wary and conscious, all the more so that he has at his disposal presently all the means needed for timely forecasting and prevention of such dangers.

The above suggests immediately two implications (Drobyshevski *et al* 2009) which thus far, as far as we are aware, have not been raised in connection with the TP, because its problem was considered from a totally different standpoint:

(1) When will this unexploded part of the cometary nucleus return to the Earth?

(2) Is there among near-Earth minor bodies an object that approached the Earth on June 30, 1908, and, if there is, what was its trajectory evolution before and after this date?

We now see that the TP problem is related to not a small extent with the problem of the so-called asteroid/comet danger for the Earth. It appears noteworthy in this connection to recall one more implication of the NEC, namely, the potential danger for Mankind of explosion of the icy envelope of Callisto, the fourth Galilean satellite of Jupiter (Drobyshevski 1989, 2008b) and the ensuing need for organization of a space mission with the purpose of probing the extent of saturation of Callisto's ices by products of their electrolysis. Speaking by Cato the Elder (234-149 BC) words, "In my opinion, Callisto must be explored to prevent its possible explosion".

---

[7] The second scenario is a fall into atmosphere of a dense cloud of interplanetary dust.

**Acknowledgments**

The author expresses his gratitude to M.E. Drobyshevski for posing the far-reaching question of the possibility for the unexploded part of the nucleus of the P/Tunguska-1908 comet to return to the Earth. Thanks are due also to J.R. Marlon for elucidative discussions.